# A new supervised non-linear mapping

Sylvain Lespinats, Anke Meyer-Baese, Michaël Aupetit

S. Lespinats is with the CEA/INES, Laboratory for Solar Systems (L2S), BP 332, 50 avenue du Lac Léman, F-73377 Le-Bourget-du-lac, France. E-mail: sylvain.lespinats@cea.fr.
A. Meyer-Baese is with Department of Electrical and Computer Engineering, Florida State University, Tallahassee, FL 32310-6046 , USA. E-mail: amb@eng.fsu.edu
M. Aupetit is with the CEA LIST, Information, Models and Machine Learning Laboratory (LIMA), F-91191 Gif-sur-Yvette, France. E-mail: michael.aupetit @cea.fr.

Abstract— Supervised mapping methods project multi-dimensional labeled data onto a 2-dimensional space attempting to preserve both data similarities and topology of classes. Supervised mappings are expected to help the user to understand the underlying original class structure and to classify new data visually. Several methods have been designed to achieve supervised mapping, but many of them modify original distances prior to the mapping so that original data similarities are corrupted and even overlapping classes tend to be separated onto the map ignoring their original topology. We propose ClassiMap, an alternative method for supervised mapping. Mappings come with distortions which can be split between tears (close points mapped far apart) and false neighborhoods (points far apart mapped as neighbors). Some mapping methods favor the former while others favor the latter. ClassiMap switches between such mapping methods so that tears tend to appear between classes and false neighborhood within classes, better preserving classes' topology. We also propose two new objective criteria instead of the usual subjective visual inspection to perform fair comparisons of supervised mapping methods. ClassiMap appears to be the best supervised mapping method according to these criteria in our experiments on synthetic and real datasets.

1. Context
Non-linear mapping is a set of powerful tools often used for visual analysis of datasets. High dimensional data are mapped as points in a low-dimensional vector space while preserving "as much as possible" dissimilarities or distances between data points. The present article focuses on the mapping of labeled data, the so called ClassiMap.

2. Classimap: a new supervised non-linear multidimensional scaling method which does not modify original distances

As in other supervised methods, we need to distinguish between pairs of data points with the same or different labels. However, we need to keep original distances unchanged. To make this possible, we propose to use the weighting function that usually emphasizes small distances in stress-driven methods. Indeed, we can take advantage of the well-known features of Sammon's mapping [1] and Curvilinear Component analysis (CCA) [2] : minimizing the stress of the former favors false neighborhoods while minimizing the stress of the latter favors tears [3].
ClassiMap stress function simply switches between Sammon and CCA stress functions based on the class label of pairs of data points, so that tears are favored between classes and false neighborhood within classes.

We define A the co-membership of data points to classes. A is a square matrix n×n where n is the number of data points. Just as in a distance matrix, the item Aij located in the ith line and the jth column corresponds to the class relationship between data points i and j. We set Aij = 1 if the data points i and j belong to the same class, and Aij = 0 otherwise.
The ClassiMap' local stress is written:
$$E_{Classimap}(i,j) = \left| d_{ij} - d_{ij}^* \right|^p \times \left( A_{ij} \times F(d_{ij}) + (1 - A_{ij}) \times F(d_{ij}^*) \right)$$

Just as Sammon's and CCA' stresses, the ClassiMap stress is equal to 0 when the map is perfect. Therefore we can expect to end up with a map having the least distortion with respect to the original pairwise distances exactly as for unsupervised mappings. However, unavoidable distortions will be driven through tears between classes and false neighborhoods within classes, thus where they maximally preserve the original topology of the classes and the readability of the map for subsequent visual classification.

Many functions can be used for F. In the following, we use
$$F(x) = 1 - \int_{-\infty}^{x} f(u, \mu, \theta) du$$

where
$$\mu = \operatorname*{mean}_{i,j}(d_{ij}) - 2 \times (1-\lambda) \times \operatorname*{std}_{i,j}(d_{ij}),$$
$$\theta = -2 \times \lambda \times \operatorname*{std}_{i,j}(d_{ij}),$$
and $f(u, \mu, \theta)$ is the probability density function of a Gaussian variable with mean µ and standard deviation (std) θ, as proposed in DD-HDS algorithm [3]. The parameter λ controls the weight of the neighborhood. It decreases (linearly) during the course of the algorithm. In the presented maps, λ is at first set to 0.9 and its final value is 0.1. The parameter p is set to 1.

ClassiMap is a multidimensional scaling method, the input of the algorithm is a distance or dissimilarity matrix with no need for a Euclidean hypothesis.

[1] Sammon J.W., A nonlinear mapping for data structure analysis, IEEE Trans. Comput., vol. C-18, no. 5, pp. 401–409, May 1969.
[2] Demartines P. and Hérault J., Curvilinear component analysis: A selforganizing neural network for nonlinear mapping of data sets, IEEE Trans. Neural Netw., vol. 8, no. 1, pp. 148–154, Jan. 1997.
[3] Lespinats S., Verleysen M., Giron A., Fertil B., DD-HDS: a tool for visualization and exploration of highdimensional data, IEEE Trans. Neural Netw., vol. 18, no. 5, pp. 1265-1279, 2007.